\documentclass[aps,prl,twocolumn,showpacs,floatfix]{revtex4-1}
\usepackage{graphicx}
\usepackage{amsmath}
\usepackage{amssymb}
\usepackage{url}
\usepackage{xcolor}

\newcommand{\comment}[1]{{\textcolor{black}{#1}}}

\graphicspath{{.}{../}{./eps/}}

\begin{document}
\title{Metal-to-insulator transition and electron-hole puddle formation\\
  in disordered graphene nanoribbons
}

\author{Gerald Schubert}
\author{Holger Fehske}
\affiliation{Institut f\"ur Physik, Ernst-Moritz-Arndt-Universit\"at
  Greifswald, 17487 Greifswald, Germany}

\date{\today}

\begin{abstract}

  The experimentally observed metal-to-insulator transition in
  hydrogenated graphene is numerically confirmed for actual sized
  graphene samples and realistic impurity concentrations.
  The eigenstates of our tight-binding model with substitutional
  disorder corroborate the formation of electron-hole-puddles with
  characteristic length scales comparable to the ones found in
  experiments.
  \comment{The puddles cause charge inhomogeneities and tend
    to suppress Anderson localization.}
  Even though, monitoring 
  the charge carrier quantum dynamics and performing a finite-size scaling
  of the local density of states distribution, we find strong
  evidence for the existence of localized states in graphene nanoribbons with 
  short-range but also correlated long-range disorder.  
\end{abstract}

%\pacs{71.23.An}{Theories and models; localized states}
%\pacs{72.15.Rn}{Localization effects (Anderson or weak localization)}
%\pacs{71.30.+h}{Metal-insulator transitions and other electronic
%transitions}
%\pacs{05.60.Gg}{Quantum transport}

%\pacs{72.20.Ee}{Mobility edges; hopping transport}

\maketitle

%%%%%%%%%%%%%%%%%%%%%%%%%%%%%%%%%%%%%%%%%%
%\section{Introduction}
%%%%%%%%%%%%%%%%%%%%%%%%%%%%%%%%%%%%%%%%%%

The experimental observation of a disorder-induced metal-to-insulator
transition in graphene upon hydrogenation~\cite{BMESHKR09} has
triggered a vivid debate on the nature of this transition.
\comment{For high concentrations of hydrogen several mechanisms of gap
  opening have been discussed, such as full $sp^2$ to $sp^3$
  transition, localization of $sp^3$ areas, or erasing of midgap
  states ~\cite{ENMMBHFBKGN09}~\footnote{Note that we focus
    exclusively on transport properties and do not address other
    interesting issues such as defect-induced magnetism or enhanced
    chemical activity~(D. W. Boukhvalov and M. I. Katsnelson, 
    J. Phys. Chem. C {\bf 113}, 14176 (2009); O. V. Yazyev, Rep. 
    Prog. Phys {\bf 73}, 056501 (2010))}}.
Graphene, on the other hand, is a truly two-dimensional system, and
the one-parameter scaling theory predicts that at zero temperature any
finite amount of disorder should lead to Anderson localization
(AL)~\cite{AALR79}.
Otherwise, the existence of a scaling function might be questionable
since the Fermi wavelength diverges near the charge neutrality point
and there is no spatial scale on which the conductivity is much larger
than $e^2/h$~\cite{NM07}.
% 
%%%%%%%%%%%%%%%%%%%%%%%%%%%%%%%%%%%%%%%%%%%%%%%%%%%%%
% experimental absence due to e-h-puddles           %
%%%%%%%%%%%%%%%%%%%%%%%%%%%%%%%%%%%%%%%%%%%%%%%%%%%%%
%
So far it seems that AL has not been seen in disordered graphene down
to temperatures of liquid-helium~\cite{CGPNG09}.
This surprising result has been attributed to strong charge carrier
density fluctuations that break up the sample into electron-hole
puddles~\cite{ACFD08}.
Within these puddles the local chemical potential deviates enough from
the charge neutrality point to allow for electron or hole
conductivity.
\comment{Mesoscopic transport is then determined by activated
  (variable-range) hopping or leakage between the
  puddles~\cite{DAHR11}.
  If the formation of electron-hole puddles is suppressed, however, AL
  might be observed. This has been reported by quite recent experiments in
  double-layer graphene heterostructures}~\cite{PZJMNCFWTGG11p}.
%
% \mycreplace{}{
%   It should be noted that the absence of diffusion due to 
%   AL fundamentally differs from the effect
%   of a gap opening upon hydrogenation~\cite{ENMMBHFBKGN09}. 
%   %An in-depth discussion of the latter beyond the scope of this work.
%   Discussing the latter is beyond the scope of this work.}

%%%%%%%%%%%%%%%%%%%%%%%%%%%%%%%%%%%%%%%%%%%%
%  LRI vs. SRI for Dirac and tight-binding %
%%%%%%%%%%%%%%%%%%%%%%%%%%%%%%%%%%%%%%%%%%%%
%
Previous theoretical work on disordered graphene strongly emphasizes the
difference between short- and long-range scatterers.
While the former applies to the case of hydrogenation, the latter
rather describes the effect of charged impurities in the
substrate~\cite{KCTSMI10}.
Within the Dirac approximation, only short-range impurities cause
intervalley scattering, and thus may lead to AL~\cite{SA02}.
The presence of long-range impurities alone gives rise to intravalley
scattering which is not sufficient to localize the charge
carriers~\cite{BTBB07,*NKR07}.
Another factor is the edge geometry of the graphene nanoribbons (GNRs)
that determines the universality class of disordered samples as long
as the phase coherence length exceeds the system size~\cite{WTS07}.
Going beyond the Dirac approximation and describing graphene by a
tight-binding model, it is natural to ask whether the scattering range
is still decisive.
Deviations from the idealized linear dispersion, a finite lattice
spacing, and the trigonal lattice symmetry, which breaks the
rotational symmetry of the Dirac cones, call for a careful numerical
analysis of the localization properties within the tight-binding
description~\cite{XX07,*BC10,*CCRFP10}.

%%%%%%%%%%%%%%%%%%%%%%%%%%%%%%%%%%%%%%%%%%%%%%%%%%
%  In this work...                               %
%%%%%%%%%%%%%%%%%%%%%%%%%%%%%%%%%%%%%%%%%%%%%%%%%%
%
In this work, we prove by unbiased numerics that experimentally
relevant concentrations of hydrogen \mbox{$x\ll 1$} may induce a
metal-to-insulator transition in actual-size graphene samples.
We show that the single-particle wavefunctions of our disorder
model are also localized for correlated long-range disorder.
Even for potential fluctuations on an atomistic scale there is
strong evidence for electron-hole puddle formation on an intrinsic
scale of some ten nm, in agreement with recent experimental
observations.~\cite{DBMLL09,*ZBGZC09,*XSBJDWTJL11}.
In contrast to previous studies on temperature dependent transport in
disordered graphene within the semiclassical Boltzmann
approach~\cite{DHL11p}, we restrict ourselves in the following to
strictly zero temperature and adopt a purely quantum point of view.

% %%%%%%%%%%%%%%%%%%%%%%%%%%%%%%%%%%%%%%%%%%
% \section{Model and method}
% %%%%%%%%%%%%%%%%%%%%%%%%%%%%%%%%%%%%%%%%%%

\begin{figure}\centering
  \includegraphics[width=0.8\linewidth,clip]{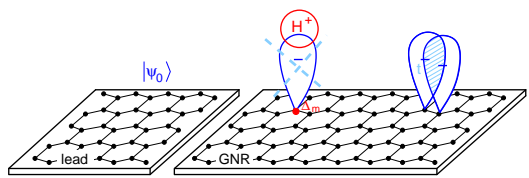}
  \caption{(Color online) Cartoon of the substitutional disorder model
    describing hydrogenated graphene.}
\label{fig:Sketch}
\end{figure}

We consider a tight-binding Hamiltonian
%
%\begin{equation}
%  \label{model}
$  {H} =
- t \sum_{\langle ij \rangle}
\bigl({c}_i^{\dag} {c}_j^{} + \text{H.c.}\bigr) + 
\sum_{i} V_i {c}_i^{\dag} {c}_i^{} ,
$
%\end{equation}
%
on the honeycomb lattice with $N$ sites, where the
operators ${c}_i^{\dag}$ (${c}_i^{}$) create (annihilate) an electron
in a Wannier state centered at site $i$, and $t$ denotes the
nearest-neighbor transfer integral.
The landscape of onsite potentials $\{V_i\}$ results from the
superposition of contributions of $N_{\text{imp}}=xN$ randomly
distributed Gaussian impurities at positions ${\mathbf r}_m$~\cite{RTB07}:
%\begin{equation}
$  V_i = \sum_{m=1}^{N_{\text{imp}}} \Delta_m 
  \exp\left(-|{\mathbf r}_i-{\mathbf r}_m|^2/(2\xi^2)\right)\,.
$
%\end{equation}
%
By choice of $\xi$ the range of the individual impurity potentials can be
continuously tuned from short-ranged to long-ranged.
For $\xi\to0$ and $x=1$ we recover the Anderson model on
a GNR~\cite{ssf09}.
Assuming a fixed $\Delta_m=\Delta$ for all impurities, the limit
$\xi\to0$ results in the binary alloy model in which 
only distinct sites acquire a finite onsite potential.
Vacancies correspond to sites with $\Delta\to \infty$, leading to a
quantum site-percolation scenario~\cite{SF08}.
The presence of adsorbed hydrogen atoms alters the hybridization of
carbon atoms from $sp^2$ to $sp^3$, partially removing the
corresponding $p_z$ orbital from the $\pi$-band.
We model the yet finite probability of finding electrons at the
adsorbant site by a finite value of the disorder strength $\Delta$
(see Fig.~\ref{fig:Sketch}).
%

%%%%%%%%%%%%%%%%%%%%%%%%%%%%%%%%%%%%%%%%%%
%\section{Results and Discussion}
%%%%%%%%%%%%%%%%%%%%%%%%%%%%%%%%%%%%%%%%%%

\begin{figure}\centering
  \includegraphics[width=0.9\linewidth,clip]{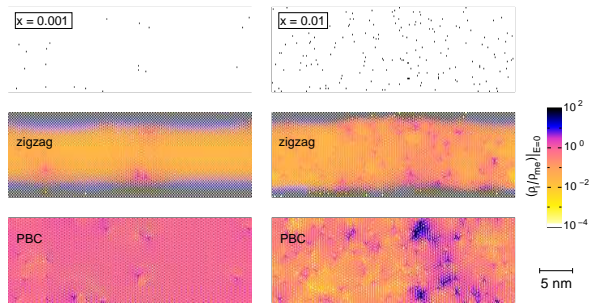}
  \caption{(Color online) Spatial distribution of the normalized LDOS
    $\rho_i/\rho_{\text{me}}$ at $E=0$ for the binary alloy model with
    potential difference $\Delta=6.0t$ and impurity concentration
    $x=0.1\%$ (left column) and $1\%$ (right column).
    The disorder configurations shown in the top panel were used for
    both BC.  Data obtained by exact diagonalization
    (ED), GNR sample size $(37\!\times\!12)\,\text{nm}^2$, corresponding to
    $300\!\times\!60$ atoms.}
  \label{LDOS_snapshot_bc}
\end{figure}
Experimental results by Bostwick {\it et al.}~\cite{BMESHKR09} suggest a
metal-to-insulator transition in graphene for a hydrogen coverage as
low as $0.3\%$.
In Fig.~\ref{LDOS_snapshot_bc} we contrast the spatial distribution of
the local density of states (LDOS)
%\begin{equation}
 $ \rho_i(E) = \sum_n |\langle n|i\rangle|^2\delta(E-E_n)$
%\end{equation}
at the Dirac point energy for a hydrogen coverage slightly below and
above this threshold.
For zigzag boundaries the well known edge states persist even in the
presence of weak disorder.
Impurities on the sublattice with high LDOS values near the GNR edges
drastically reduce these values.
On the other sublattice they do not have any effect.
In the bulk of the ribbon, the LDOS is slightly enhanced as compared
to the ordered case.
Positive interference traps the wavefunction on sites in between the
impurities.
For periodic boundary conditions (PBC) the spatial LDOS distribution
is clearly distinct for both impurity concentrations:
We observe only slight local perturbations of the perfectly extended
state for low impurity concentrations but a clearly localized state at
$1\%$ hydrogen coverage.
Measurements on a sample of this size therefore yield metallic
(insulating) behavior for coverages $0.1\%$ ($1\%$).
The observed metallic character seems to be merely a consequence of a
finite localization length $\lambda$ that exceeds the system size.
Note that the calculated state characteristics and experimental
results agree qualitatively for PBC only.
This underlines that the observed localization properties are
intrinsic to short-range disordered bulk graphene.
Edge effects arise on top, but are to a certain extent irrelevant in
experiments, especially if mobilities are measured using a
multiterminal Hall geometry~\cite{KSZXH09}.

\begin{figure}\centering
  \includegraphics[width=0.9\linewidth,clip]{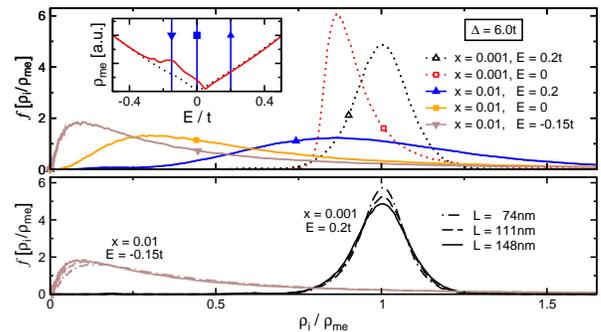}
  \caption{(Color online) \comment{
      Upper panel: Distribution of the
      LDOS at experimentally relevant energies for the
      binary alloy model with different impurity concentrations and
      PBC.  The sample width $W=109\,\text{nm}$. Normalization of the
      distribution to $\rho _{\text {me}}$ directly relates its
      position to the height of the maximum and its width. Inset:
      Magnification of the averaged DOS with indications of the
      energies for which the LDOS distributions are shown. Lower
      panel: Finite-size scaling of the LDOS distribution. Data
      obtained by the kernel polynomial method with resolution adapted
      to the level-spacing, $N_k=140$ (for details
      see~\cite{WWAF06,ssbfv10}).  } }
\label{LDOS_Binalloy_perc}
\end{figure}
\comment {In order to assert that AL takes place in infinite GNRs, we
  analyze the distribution of the normalized LDOS in
  Fig.~\ref{LDOS_Binalloy_perc}.}
We restrict ourselves to three characteristic energies which are shown
in the inset together with the averaged density of states
$\rho_{\text{me}}=\langle\rho_i\rangle$.
The behavior of the LDOS distribution upon finite-size scaling 
(lower panel) is a powerful criterion to detect AL for 
different kinds of disordered systems~\cite{ssbfv10}, even in presence of
interactions~\cite{BAF04,*SBH11}.
Extended states are characterized by an $f[\rho_i/\rho_{\text{me}}]$
being independent of the system size.
Otherwise sensitivity of the distribution to the system size indicates
localization, which we indeed observe for binary alloy disordered GNRs
for all energies and both impurity concentrations.

\comment{For a given state, the shape of the LDOS distribution
and the extent of its shifting depend
on $\lambda$; the more 
pronounced the shift and the more asymmetric $f[\rho_i/\rho_{\text{me}}]$,
the shorter is $\lambda$.
Larger impurity concentrations enhance localization, 
as can be seen from the asymmetric shape of $f[\rho_i/\rho_{\text{me}}]$ 
for $x=0.01$.
The persisting size dependence for $x=0.001$
proves localization also for such a weak randomness.
Here the high sensibility of the LDOS distribution to the ratio of 
$\lambda$ and system size is of vital importance.
It allows to detect localization also in the case of weak
disorder for which $\lambda$ distinctly exceeds the
system size and consequently $f[\rho_i/\rho_{\text{me}}]$ is
concentrated around unity.}

\begin{figure}\centering
  \includegraphics[width=\linewidth,clip]{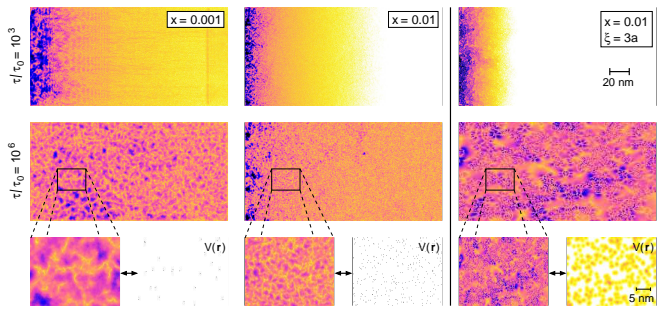}
  \caption{(Color online) Time evolution of the local particle density
    on disordered zigzag GNRs after,  at $\tau=0$,  finite (ordered)
    leads were attached to the left and right of the GNR.      
    $|\psi(\mathbf{r}_i,t)|^2$ is
    normalized to the actual mean particle density on the GNR.  The
    color scale is identical to Fig.~\ref{LDOS_snapshot_bc}. Sample
    dimensions are $(221\times109)\,\text{nm}^2$, corresponding to
    $1800 \times 512$ lattice sites. Left and
    middle column: Binary alloy model with $\Delta=6t$. Right column:
    Gaussian correlated disorder model with $\xi=3a$, where $a$ is the
    inter-carbon distance. Here the potential is normalized to
    $\max(V_i)=6t$. The insets show magnifications of
    $|\psi(\mathbf{r}_i,t)|^2$ in the quasistationary regime together
    with the corresponding potential landscape.} 
\label{fig:TimeEv}
\end{figure}
\begin{figure}\centering
  \includegraphics[width=0.9\linewidth,clip]{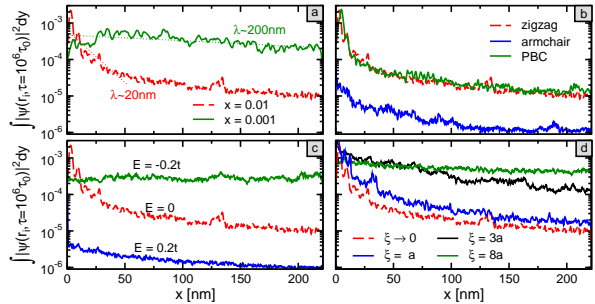}
  \caption{(Color online) 
    \comment{Quasistationary local particle
      density summed over the transverse direction. Each subpanel
      shows how changing one control parameter influences the baseline 
      case (red dashed line: binary disorder, $x=0.01$, $E=0$,
      zigzag BC) while keeping the others fixed.}
  }
  \label{fig:TimeEv_quantitative}
\end{figure}
In Fig.~\ref{fig:TimeEv} we contrast the quantum dynamics of a
particle injected into zigzag GNRs with binary or Gaussian correlated
disorder and impurity concentrations of $0.1\%$ and $1\%$.
As initial state $|\psi_0\rangle$ we prepare an exact $E=0$ eigenstate
of the ordered infinite graphene lattice in the lead left to the sample
(see Fig.~\ref{fig:Sketch}).
After bringing the lead in contact with the sample, we let the system
evolve in time by solving the time-dependent Schr{\"o}dinger equation
by a Chebyshev expansion technique~\cite{SF08,FSSWFB09,*LSYD11}.
Due to the coupling with the disordered GNR, $|\psi_0\rangle$ is not an
eigenstate of the overall system but comprises admixtures of other
states, mainly from the vicinity of $E=0$.
The snapshot at $\tau=10^3\tau_0$, where $\tau_0=\hbar/t$, confirms
the intuition that spreading is faster the lower the impurity
concentration is (see Fig.~\ref{fig:TimeEv}).
At $\tau=10^6\tau_0$ all states have reached quasistationarity, and we
can extract their characteristic features.
For $x=0.1\%$ the state spans the whole sample.
The inset reveals its puddle-like structure with density fluctuations
of two orders of magnitude on length scales of $5-10\,\text{nm}$.
At an impurity concentration of $x=1\%$ the particle density is
reduced about two orders of magnitude between left and right edge of
the GNR, providing a direct visualization of AL.
Note that the local structure of $|\psi\rangle$ remains puddle-like,
but the spatial extent of the puddles is substantially reduced below 
$1\,\text{nm}$.
From the similarity of both local structures one may argue that for
larger systems also states for $x=0.1\%$ will be localized.
Note that correlated disorder results in a markedly smoother potential
landscape.
\comment{
In addition, the electron-hole puddles are superimposed by
a coarse-grained filamentary structure.
}

\comment{
Integrating the local particle density over the transverse ribbon
direction allows for a more quantitative analysis of the localization
properties (see
Fig.~\ref{fig:TimeEv_quantitative}).
For a fixed configuration of impurity positions we vary the relevant
control parameters and extract $\lambda$ from fitting
the  quasistationary density to an exponential decay.
Removing part of the impurities results in a larger $\lambda$
[Fig.~\ref{fig:TimeEv_quantitative}(a)].
For the considered ribbon width the difference between zigzag- and
PBC is marginal, while armchair edges
drastically reduce the transmission
[Fig.~\ref{fig:TimeEv_quantitative}(b)].
This reflects the mismatch of preferred transport direction and ribbon
axis.
Varying the incident particle energy 
[Fig.~\ref{fig:TimeEv_quantitative}(c)]
we observe an enhanced transmission for $E=-0.2t$, which might
be attributed to resonant (localized) states. 
Note that at the position of the chemical potential of graphene on a SiC substrate,
$E=0.2t$, $\lambda$ is even shorter than 
at the Dirac point $E=0$.
For correlated long-range disorder
[Fig.~\ref{fig:TimeEv_quantitative}(d)], $\lambda$
increases with $\xi$, yielding extended states if $\xi\gg a$.
For $\xi=a,3a$ the states are still clearly localized.
This disagrees with results in the literature obtained within the
Dirac approximation~\cite{BTBB07,*NKR07}, where the valley quantum
number is conserved and localization is suppressed in absence of
intervalley scattering~\cite{SA02}.
}
We attribute this difference to the lattice discreteness and the
breaking of the rotational symmetry of the Dirac cones by the
trigonal symmetry of the honeycomb lattice.
Moreover, even for a narrow banded initial state the dynamics is
influenced by states from the whole energy spectrum where the graphene
dispersion significantly deviates from the linear approximation.
If all these aspects were taken into account, long-ranged disorder may
cause localization within a tight-binding description in accordance
with one-parameter scaling~\cite{ZHBWXL09}.
\begin{figure}\centering
  \includegraphics[width=0.9\linewidth,clip]{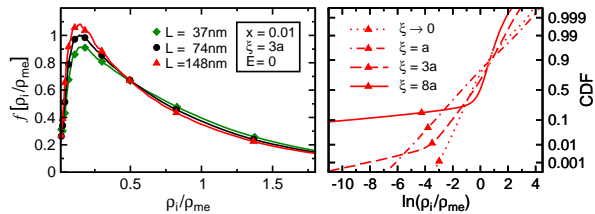}
  \caption{(Color online) Left: Finite-size scaling of the
      LDOS distributions for the Gaussian correlated disorder model
      with parameters matching Fig.~\ref{fig:TimeEv}. Right: CDF of
      the LDOS for various potential ranges $\xi$.
    }
\label{LDOS_LongRangeDisorder}
\end{figure}
Whether AL really occurs for correlated disorder can be proven by
performing a finite-size scaling of the LDOS distribution.
In doing so we find evidence for localization for
  $\xi=3a$ from the shifting of the LDOS distribution (see left panel of
Fig.~\ref{LDOS_LongRangeDisorder}).
A remarkable feature of $f[\rho_i/\rho_{\text{me}}]$ is the salient tail that
develops for small values of $\rho_i$ on increasing $\xi$.
In the right panel of Fig.~\ref{LDOS_LongRangeDisorder} this results
in a kinked cumulated distribution function (CDF) instead of the
approximate straight CDF for uncorrelated disorder~\footnote{ We would
  like to stress that there is numerical evidence against a strict
  log-normal shape of the LDOS distribution,%~\cite{RVSR11},
  for which the CDF would be an exact straight line, see Rodriguez 
  \it{et al.} (2009), arXiv:1107.5736v1.}.
 
\begin{figure}\centering
  \includegraphics[width=\linewidth,clip]{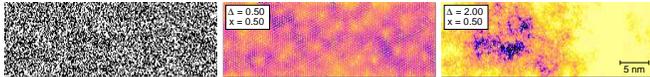}
  \caption{(Color online) Spatial distribution of the normalized LDOS
    at $E=0$ for the disorder configuration shown in the left
    panel
    with PBC. System sizes and color mapping are the same as in
    Fig.~\ref{LDOS_snapshot_bc}.  Data obtained by ED.}
  \label{snapshot_tunnel}
\end{figure}

So far we considered strong disorder ($\Delta\gg1$) and weak
randomness ($x\ll1$), for which it is tempting to relate average
puddle size and distance between the impurities.
Interestingly, electron-hole puddles also arise for weak disorder
strength and strong randomness, corresponding to a disorder landscape
varying on an atomistic scale without any correlations.
Such a modeling might be regarded as an attempt to capture the effect
of the buffer layer forming between epitaxially grown graphene and its
SiC substrate~\cite{EBHJKLMORRRSWWS09}.
In this setup we refrain from considering a correlated potential
landscape in order not to impose any a priori correlations in the
LDOS.
For weak disorder strength, $\Delta=0.5t$, the LDOS at $E=0$
nevertheless becomes puddle-like with a characteristic scale of
$2$-$5\,\text{nm}$ (see middle panel of Fig.~\ref{snapshot_tunnel}).
Note that the choice of BC has no qualitative impact
on the LDOS in this limit of the binary alloy model.
Obviously, any subtle differences in the state characteristics induced
by BC are masked by the randomness of the potential
landscape.
Increasing the potential difference to $\Delta=2t$, $\lambda$ drops
below the system size and we observe clearly localized states.
%

%%%%%%%%%%%%%%%%%%%%%%%%%%%%%%%%%%%%%%%%%%
%\section{Conclusions}
%%%%%%%%%%%%%%%%%%%%%%%%%%%%%%%%%%%%%%%%%%
\comment{ To conclude, hydrogenated graphene behaves at zero
  temperature as a 'normal' two-dimensional disordered system
  concerning AL, provided the extensions of ultra-high-quality samples
  become very large.
  If the localization length noticeably exceeds the system size, the
  sample nevertheless shows metallic behavior.
  We find that also certain long-range correlations in the disorder
  landscape yield localized single-particle wavefunctions.
  Most notably, we show that disorder-induced electron-hole puddles
  may arise for both disorder types.
  The intrinsic scale of the puddle-like
  structures in the eigenstates is not simply set by the distance
  between impurities, but results from subtle quantum interference
  effects.
  Even for atomic scale fluctuations of the disorder potential they
  might exceed 5 nm, which is in the range of experimentally measured
  values.
  The presence of electron-hole puddles, leading to intra- and
  inter-puddle transport, drives the system away from the
  metal-to-insulator transition, thereby masking
  AL~\cite{PZJMNCFWTGG11p}. 
  This resolves the puzzle why AL is so hard to detect in disordered
  graphene and GNRs.
}

We acknowledge financial support by DFG through the graphene priority
program SPP 1459, KONWIHR-II as well as granting of computing time on
HLRB Munich.
% and Regional Computing Center Erlangen.

%\bibliography{./ref}
%Merlin.mbs v4.21 2009-07-09.

%

\end{document}